\documentclass[pre,aps,
amssymb,amsmath,superscriptaddress,showpacs,showkeys]{revtex4}
\usepackage{graphicx}

\newcommand{\ZM}{{\mathbb Z}}

\newcommand{\Hop}{{\hat H}}
\newcommand{\Hopr}{{\hat H}_{r}}
\newcommand{\Poprr}{{\hat P}_{\mbox{\tiny R}}}
\newcommand{\Pop}{{\hat P}}

\newcommand{\Vop}{{\hat V}}

\newcommand{\Uopp}{{\hat{\cal U}}}
\newcommand{\Uop}{{\hat U}}

\newcommand{\dk}{D_m}
\newcommand{\dz}{D_0}
\newcommand{\Ir}{{\mathcal I}}

\newcommand{\hres}{{\overline H}_{\mbox{\tiny res}}} 

\begin{document}

\title{Decay of Quantum Accelerator Modes.}
\author{Michael Sheinman}
\affiliation{Physics Department,  Technion, Haifa 32000, Israel.}
\author{Shmuel Fishman}
\affiliation{Physics Department,  Technion, Haifa 32000, Israel.}
\author{Italo Guarneri},
\affiliation{Center for Nonlinear and Complex Systems, Universit\'a dell'Insubria,
Via Valleggio 11, I-22100 Como, Italy.}

\affiliation{Istituto Nazionale di Fisica Nucleare, Sezione di Pavia,
via Bassi 6, I-27100 Pavia, Italy.}

\affiliation{CNISM - Sezione di Como - via Valleggio 11, 22100 Como, Italy.}
\author{Laura Rebuzzini}
\affiliation{Center for Nonlinear and Complex Systems, Universit\'a dell'Insubria,
Via Valleggio 11, I-22100 Como, Italy.}

\affiliation{Istituto Nazionale di Fisica Nucleare, Sezione di Pavia,
via Bassi 6, I-27100 Pavia, Italy.}

\pacs {05.45.Mt, 03.75.-b, 42.50.Vk}

\keywords {tunneling, cold atom optics, phase space structures}

\begin{abstract}
{Experimentally observable Quantum Accelerator Modes are used as a
test case for the study of some general aspects of  quantum decay
from classical stable islands immersed in a chaotic sea.  The modes
are shown to correspond to  metastable states, analogous to the
Wannier-Stark resonances. Different regimes of tunneling, marked by
different quantitative dependence of  the lifetimes on $1/\hbar$,
are identified, depending on the resolution of KAM substructures
that is achieved on the scale of $\hbar$. The theory of Resonance
Assisted Tunneling introduced by Brodier, Schlagheck, and Ullmo
\cite{BSU02}, is revisited, and found to well describe decay
whenever applicable.}
\end{abstract}

\maketitle

\section{Introduction.}

\noindent Classical Hamiltonian systems generically display mixed
phase spaces, where regions of regular motion and regions of chaotic
motion coexist \cite{LL92}. Although no classical transport is
allowed between different, disjoint regions,  quantum transport is
made possible by so-called "dynamical tunneling", which allows
wave-packets to leak through classical invariant curves. In the case
of a wave-packet initially localized inside a classical stable
island, a transfer of probability into the chaotic region arises,
that may continue a long time, and thus take the form of
irreversible decay from the island, whenever $\hbar$ is so small
that the Heisenberg time related to fast chaotic diffusion is much
larger than the period(s) of regular motion in the island. Islands
of regular motion have been found to play crucial roles in many
contexts, such as optical cavities
\cite{oc}, driven cold atoms \cite{aop}, where
they have inspired detailed theoretical analysis \cite{nimrod}, and
 billiards , where they motivated
mathematical investigations \cite{vered} and atom-optical experimental realizations
\cite{davidson}.
Wide attention has been
attracted by Chaos Assisted Tunneling, which denotes  transmission between
symmetry-related islands, across a chaotic region \cite{cat}.
Slow tunneling out of islands
dominates the dynamical localization properties in some
extended systems \cite{iomin,BKM05}. \\
In this paper we explore decay from regular islands into the
surrounding chaotic regions, which is a central theoretical issue
in all the above hinted subjects. It is an appealing idea that decay
rates may be estimated, using only information drawn from the
structure of classical islands. On account of the ubiquitous
character  of mixed phase spaces, this theoretical programme is
attracting significant attention \cite{BSU02,NP03,BKM05,ES05}.  In
this paper we address the problem in a special case, which has a
direct experimental relevance. The Quantum Accelerator Modes
(QAM) were experimentally discovered when cold caesium atoms, falling
under the action of gravity, were periodically pulsed in time by a
standing wave of light \cite{Ox99,DSFG}. The theory of this
phenomenon \cite{FGR00} shows that the dynamics of an atom is
described, in an appropriate gauge, by a formal quantization \footnote{The role of Planck's constant
in this quantization process is not played by $\hbar$ (the Planck's
constant proper), but by a different physical parameter, which is
not, in fact, denoted $\hbar$ in the relevant papers. The theory
derives map (\ref{map}) from the Schr\"odinger equation in the limit
when this parameter tends to $0$, by a process which is
mathematically (though not physically) equivalent to taking a
classical limit. Therefore that parameter is here denoted by
$\hbar$, because its actual physical meaning is immaterial for the
theory discussed in this paper.} of
either of the classical maps :
\begin{equation}
\label{map}
{J}_{t+1}={J}_t+{\tilde k}\sin(\theta_{t+1})\pm 2\pi\Omega \;\;\;,\;\;\theta_{t+1}=\theta_t
\pm {J}_t \;\;\mbox{\rm mod}(2\pi)\;.
\end{equation}
 The classical phase portraits exhibit periodic (in
$J$) chains of regular islands. The rest of phase space is
chaotic, and QAMs are produced whenever an atomic wavepacket is at
least partially trapped inside the islands. Therefore, QAMs
eventually decay in time, due to quantum tunneling \cite{FGR00}.
The  problem of QAMs has a relation to the famous Wannier-Stark
problem \cite{GRF05} about motion of a particle under the combined action of a
constant and a periodic in space force field \cite{GKK02},  and the
class of QAMs we consider in this paper are due to metastable
states \cite{nimrod}, which are analogous to the Wannier-Stark resonances, and
are associated with sub-unitary
eigenvalues of the Floquet evolution operator. We derive and numerically support
quasi-classical,
order-of-magnitude estimates for their decay rates, based on
classical phase-space structures. We theoretically and numerically
demonstrate that decay is determined by different types of
tunneling, depending on how significant the KAM structures inside
the island are on the scale of $\hbar$, and in particular we show
that these different mechanisms result in different quantitative
dependence on the basic quasi-classical parameter given by ${\cal
A}/\hbar$, where $\cal A$ is the area of an island. In the regime
where higher-order structures inside the island are quantally
resolved, we use the theory of Resonance Assisted Tunneling
\cite{BSU02} to describe the decay of QAMs and find a good agreement
in the presence of a single dominant resonance. In particular we
observe, in especially clean form , a remarkable stepwise dependence
on the quasi-classical parameter, as first predicted in
\cite{BSU02}. \\The phase space islands which are studied in this paper are directly
related to experiments on laser cooled atoms\cite{Ox99,DSFG}.
Unfortunately the step structure which is here predicted occurs in
parameter ranges where the decay rate is extremely small. Finding
parameter ranges where decay rates are appreciable, and still
exhibit step dependence, is a challenging task for experimental
application.

\section{Classical and Quantum dynamics.}

\noindent QAMs one-to-one correspond to stable periodic orbits of
the map on the 2-torus which is obtained from (\ref{map}) on reading
$J$ modulo $(2\pi)$ \cite{GRF05}\cite{FGR00}. This class of orbits
in particular includes fixed points (i.e., period-1 orbits) of  map
(\ref{map}). It is these period-1 orbits that give rise to the most
clearly observable QAMs, and in this paper we restrict to them
\footnote {These islands are not traveling ones. They
nonetheless give rise to QAMs because map (\ref{map}) describes
motion in an accelerating frame \cite{FGR00}.}.
Map (\ref{map}) differs from the Standard Map only because of the
drift $2\pi\Omega$ in the 1st equation. Despite its formal
simplicity, this variant introduces  nontrivial problems, concerning
Hamiltonian formulation and  quantization, which are discussed in
this section.

\subsection{Wannier-Stark pendulum.}

\noindent
 Let ${\tilde k}=k\epsilon$ and $2\pi\Omega=a\epsilon$,
with $\epsilon$  a small parameter. For $\epsilon=0$, map
(\ref{map}) has circles of fixed points at the "resonant" values of
the action ${J}=2\pi s$, $(s\in\ZM)$. Straightforward calculation
shows that for $\epsilon>0$  a stable fixed point, surrounded by a
stable island, survives near each resonant action, whenever $k$ is
larger than $a$ and smaller than a stability border \cite{FGR00}.
Other islands , related to periodic orbits of higher periods, may or
may not significantly coexist with such period-1 islands, depending
on parameter values. In any case, numerical simulation shows that
motion outside all such islands is essentially chaotic at any
$\epsilon>0$. \\
Using canonical perturbation theory, one finds \cite{GRF05} that in
the vicinity of resonant actions, and at 1st order in $\epsilon$,
the dynamics (\ref{map}) are canonically conjugate to the dynamics
which are {\it locally} ruled by the "resonant Hamiltonian":
\begin{equation}
\label{ws}
 H_{res}({\cal J},\vartheta)=\frac12
{\cal J}^2+\epsilon
V(\vartheta)\;\;\;;\;\;V(\vartheta)=-a\vartheta+k\cos(\vartheta)\;.
\end{equation}
%In other words, a canonical transformation may be contrived, such
%that near the resonant actions, and in the new canonical variables
%$J,\theta$, the map dynamics differ from those which are locally
%generated by (\ref{ws}), by corrections of higher order in
%$\epsilon$.
Multivaluedness of $V(\vartheta)$ is removed on taking derivatives,
so Hamilton's equations are well-defined on the cylinder, even
though the Hamiltonian (\ref{ws}) is not. They uniquely define a
"locally Hamiltonian" flow on the cylinder, which will be termed the
Wannier-Stark (WS) pendulum, because, if $\vartheta$
 were  a linear coordinate and not an angle , then (\ref{ws})
would be the Wannier-Stark (classical) Hamiltonian  for
1-dimensional motion of a particle in a sinusoidal potential
combined with a static electric field \cite{GKK02}. Trajectories of
the Wannier-Stark pendulum are obtained, by winding around the
circle the trajectories which are defined on the line by the
Wannier-Stark
Hamiltonian. \\
 If $a=0$,
(\ref{ws}) is the Hamiltonian of a standard pendulum, and motion is
completely integrable in either of the two regions in which  phase
space is divided by the pendulum separatrix.
\begin{figure}[ht]
%  \includegraphics[width=6.3cm,angle=0]{loop.eps}
  \includegraphics[width=6.3cm,angle=0]{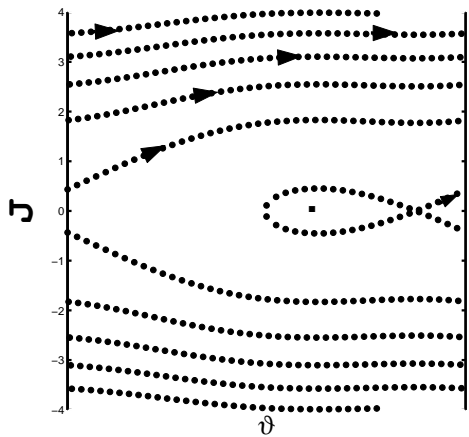}
  \caption{
Stable and unstable manifolds of a Wannier-Stark pendulum
(\ref{ws}) with $a\epsilon=0.5$ and $k\epsilon=0.7$. On the
cylinder, arrows lie on a single continuous trajectory.}\label{wssp}
\end{figure}
If $a\neq 0$ this is not true anymore. In particular,  if $|a|<|k|$
then the flow has one stable and one unstable fixed point. Motion is
completely integrable inside the stable island delimited by the
separatrix, but not outside, because trajectories are unbounded
there  and motion cannot be confined to a torus (Fig.\ref{wssp}).
The separatrix (also called a "bounce" \cite{BWH92} or "instanton
trajectory" \cite{Sch96}) is the trajectory which approaches the
unstable point, both in the infinitely far past, and in the
infinitely far future ( loop in Fig.\ref{wssp}, line PQ in
Fig.\ref{swan}).  The resonant Hamiltonian provides but a local
description of the motion near a resonance. It misses the
periodicity in action space which is an important global feature of
the problem; nevertheless, it does provide a description of the
inner structure of stable island(s).

\subsection{Quantization.}\label{quant}

\noindent Quantization of map (\ref{map}) is a nontrivial task,
because a shift in  momentum by $2\pi\Omega$, as in the 1st
eqn.(\ref{map}), may be inconsistent with quantization of momentum
in multiples of $\hbar$. This problem disappears if the angle
$\theta$ in (\ref{map}) is replaced by $x \in]-\infty,+\infty[$,
because then the  map describes motion of a particle in a line,
and straightforward quantization yields the unitary operator
$$
{\Uopp}:=e^{ia\epsilon{\hat X}/\hbar}e^{-ik{\epsilon}\cos( {\hat
X})/\hbar}e^{-i{\hat P}^2/2\hbar}\;,
$$
where ${\hat X}$ and ${\hat P}$ are the canonical position and
momentum operators. However, the quantum dynamics thus defined on
the line do not define any dynamics on the circle, except in cases
when $\Uopp$ commutes with spatial translations by $2\pi$.
This case only occurs when $a\epsilon$ is an integer multiple of
$\hbar$. Then quasi-momentum is conserved and standard Bloch theory
yields a family of well-defined rotor evolutions, parametrized by
values of the quasi-momentum. If $a\epsilon=m\hbar/n$ with $m$ and
$n$ integers, then it is easy to see that $
\Uopp^n$ commutes
with spatial translations by $2\pi$ and so the $n-$th power of the
classical map "on the circle" can be
safely quantized.\\
Similar subtleties stand in the way of quantizing the Wannier-Stark
pendulum. The WS Hamiltonian "on the line" never commutes with
translations by $2\pi$, as long as $a\epsilon\neq 0$. However, if
$a\epsilon=m\hbar/n$, then the unitary evolution generated by the WS
Hamiltonian over the integer time $n$ does commute with such
translations \cite{GKKM98}, and so it yields a family of unitary
rotor evolutions. Each of these yields a quantization
of the WS pendulum flow at such integer times. We shall
restrict to such "commensurate" cases. In the language of the theory
of Bloch oscillations \cite{GKK02}, these are the cases when the
"Bloch period" $T_B=\hbar/(a\epsilon)$ and the kicking period are
commensurate.

\section{Decay Rates.}

\noindent Let $\Uop$ generically denote the unitary operators that
are obtained by quantization of  map (\ref{map}) or powers thereof,
as discussed in Sect.\ref{quant}. We contend that, despite classical
stable islands, the spectrum of $\Uop$ is purely continuous.
Quasi-modes related to classical tori in the regular islands
correspond to metastable states, associated with
eigenvalues of $\Uop$, which lie strictly inside the unit
circle,  and thus have positive decay rates $\Gamma$ .  They are
analogous to the Wannier-Stark resonances.  Arguments supporting
this contention are presented in Appendix \ref{metastates}, along
with methods of numerically computing decay rates $\Gamma$. In this
Section we
obtain order-of-magnitude estimates of decay rates $\Gamma$.\\
Motion inside the islands is not integrable, but just
quasi-integrable, and displays  typical KAM structures, such as
chains of higher-order resonant islands. We separately consider the
cases when such structures are small (resp., large) on the scale of
$\hbar$. In the latter case, our basic theoretical tool is the
notion of Resonance Assisted Tunneling, which was introduced in
\cite{BSU02}. This theory is presented from scratch in sects.
\ref{pheno} and \ref{RAD}, with special attention to the role of
classical and quantum perturbation theory.

\subsection{Wannier-Stark tunneling}\label{wstun}

\noindent First we consider the case when $\hbar$ is small compared
to the size of an island and yet large compared to the size of the
stochastic layer and of resonant chains inside the island. This in
particular means that 2nd order corrections on the resonant
Hamiltonian (\ref{ws}) are classically small, and so one expects the
bare resonant Hamiltonian to capture the essential features.
\begin{figure}[ht]
%  \includegraphics[width=5.1cm,angle=0]{vspot.eps}
  \includegraphics[width=5.1cm,angle=0]{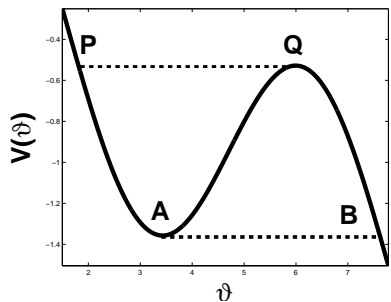}
  \caption{Wannier-Stark potential for $a=0.2$ and
$k=0.7$. }\label{swan}
\end{figure}
This Hamiltonian is formally the Wannier-Stark Hamiltonian  and
stable islands are associated with elliptic motion near the bottom
of the potential wells (Fig.\ref{swan}). A WKB estimate for the
smallest decay rate from a well is
\begin{equation}
\label{gamov} \Gamma\sim\frac{\omega_0}{2\pi}\;e^{-2S(A,B)/\hbar}
\end{equation}
where $\omega_0$ is the angular frequency of the small oscillations
and $iS(A,B)$ is the imaginary action along the classically
forbidden path from point $A$ to point $B$, at constant energy equal
to the value of the potential at the bottom of the well:
$$
S(A,B)=\int_{\theta_A}^{\theta_B}d\theta\;\sqrt{2\epsilon(V(\theta)-V(\theta_A))}\;.
$$
Reflection $\theta\to\pi-\theta$ turns $\theta_A$ into $\theta_Q$,
$\theta_B$ into $\theta_P$, and reverses the sign of the argument of
the square root; and so $S(A,B)=S(P,Q)$, the real action along the
path from $P$ to $Q$ at constant energy equal to the value of the
potential in $Q$. This path is the separatrix, so $2S(A,B)$ is equal
to the area enclosed by the separatrix, which is in turn nearly
equal to the area $\cal A$ of the actual island in the regime we are
considering. Therefore,
\begin{equation}
\label{gwsn}
\Gamma\sim\frac{\omega_0}{2\pi}e^{-{\cal A}/\hbar}.
\end{equation}
This result is compared with a numerical simulation  in
Fig.\ref{bezrez1}. Even better agreement with numerical data is
obtained by using in (\ref{gamov}) the trajectory with energy
$\hbar\omega_0/2$ above the bottom of the well, which is an
approximation to the ground state energy in the harmonic
approximation. This is shown in Figs.\ref{bezrez1} and
\ref{RezIStark1}. \\
The success of the elementary WKB approximation (\ref{gamov}) is
due to the fact that in cases like Figs.\ref{bezrez1} and
\ref{RezIStark1} the potential barriers on the right of
$\vartheta_B$ are significantly lower than $V(\vartheta_A)$, so that
tunneling trajectories have to cross just one potential barrier.
When this condition is not satisfied, one is faced with the full
complexity of the WS problem.

\begin{figure}
\includegraphics[width=10cm,angle=0]{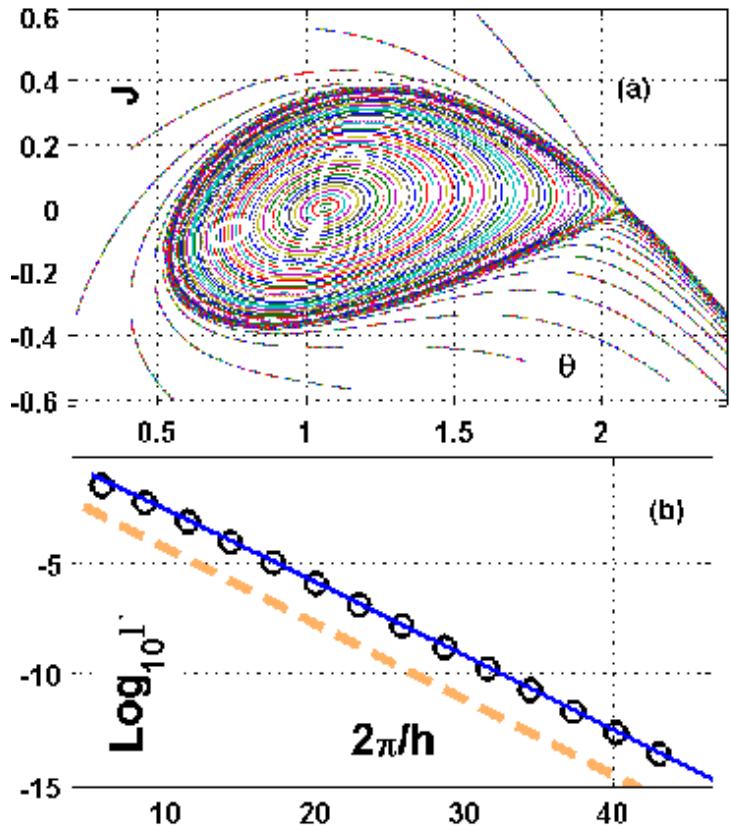}
\caption{(Color online).(a) Phase portrait of system (\ref{map}) with negative
sign, for $\tilde{k}=0.8$ and $2\protect\pi \Omega =0.7$.
% after 4000
%iterations, and  The initial conditions are $J=0$, $\theta_{Ell}
%\leq \protect\theta \leq \theta_{Hyp} $ and $\theta=\theta_{Hyp}$,
%$0 \leq \protect J \leq \protect%0.5 $
(b) Circles: numerically computed decay rate from the center of the
regular island, as a function of $1/\hbar $.
%The size of the basis
%used in the simulations  varies from 400 to 2000.
Lines show WKB estimates, obtained by using classically forbidden
paths at constant energy (\ref{ws}). These are equal either to a
minimum of the WS potential (dashed line) shown in Fig.\ref{swan},
or to the ground state energy in the potential well, estimated in
the harmonic approximation (solid line).}\label{bezrez1}
\end{figure}

\subsection{Phenomenological Quantum Hamiltonian.}\label{pheno}

\begin{figure}[ht]
%\includegraphics[width=10cm,angle=0]{RezIStark1.eps}
\includegraphics[width=10cm,angle=0]{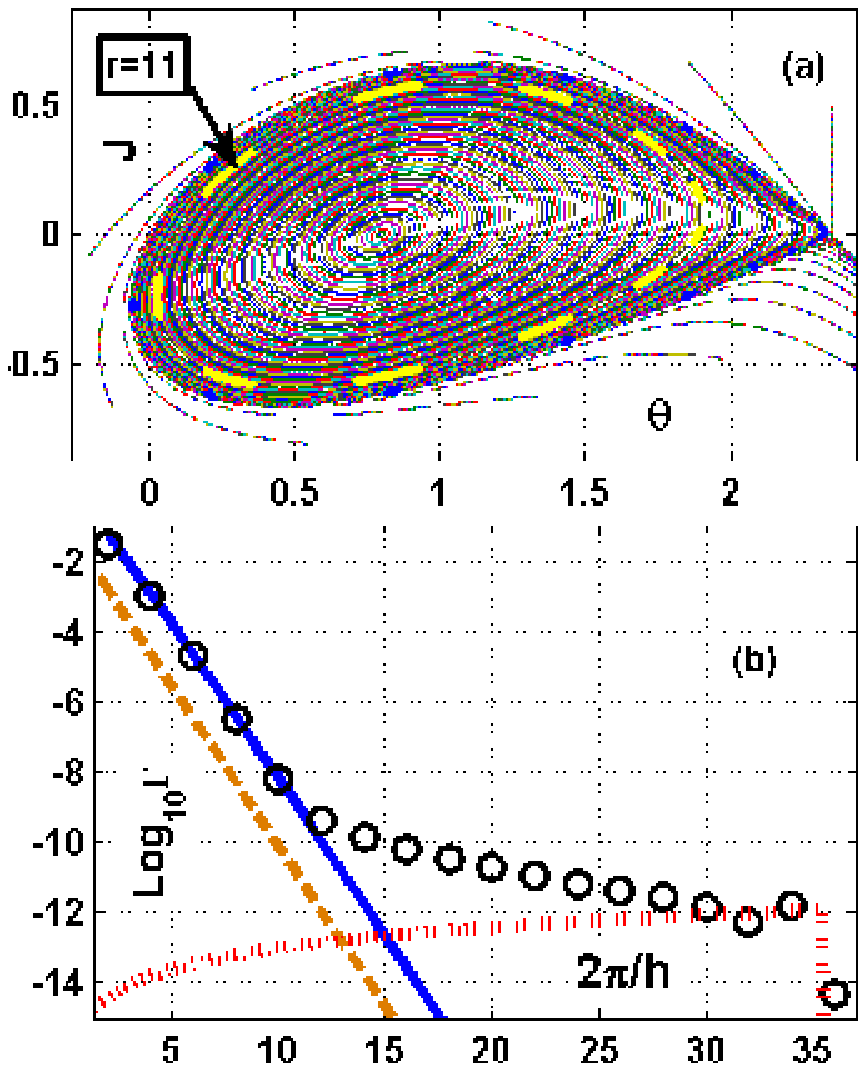}
\caption{(Color online).Same as Fig.\ref{bezrez1} for $\tilde{k}=0.7$ and $2%
\protect\pi\Omega=0.5$. The island chain highlighted in (a) is
generated by a $11:1$ resonance at $\Ir_{111}\approx0.251$, with
$v=4.376\cdot 10^{-8}$, and $M=2.785$ (see App. \ref{phspres}). The
dotted line in (b) represents the theory (\ref{ullmo}) for decay
assisted by this resonance, using the unperturbed spectrum (\ref{detu}). }\label{RezIStark1}
\end{figure}

\noindent
In Fig.\ref{RezIStark1}(b) we show the dependence of
$\Gamma$ vs $\hbar$, for the case of the island of
Fig.\ref{RezIStark1}(a). Like in the case of Fig.\ref{bezrez1}, 2nd
order resonances are not quite pronounced here, and so, at
relatively large values of $\hbar$ (leftmost part of the Figure) the
dominant contribution to decay is given by WS tunneling, and good
agreement is observed with the theory of sect.\ref{wstun}.
As $\hbar$ decreases,  a clear crossover is  observed to slower
decrease of $\Gamma$, indicating that a different mechanism of decay
is coming into play, which overrules WS tunneling. In our
interpretation, this is the mechanism of "resonance assisted
tunneling" to be discussed in what follows, which was introduced in
ref.\cite{BSU02}. It is a quantum manifestation of classical KAM
structures, and so we now consider the case when $\hbar$ is small
compared to the size of the island, yet phase-space structures
produced by higher-order corrections to the resonant Hamiltonian are
not small on the scale of $\hbar$.
The approach  to be presently described is based on a quantum Hamiltonian,
which is not formally derived from the exact dynamics, but is instead
tailored after the actual structure of the classical phase space.
This heuristic approach  is applicable when the WS linewidth (\ref{gwsn})
is negligible
with respect to the coupling between different WS tori (and
between WS tori and the continuum) which is due to higher order
corrections,  and in fact it totally misses WS tunneling given by
(\ref{gwsn}).
It is assumed  that the classical partition island/chaotic sea is
quantally mirrored by a splitting of the Hilbert space of the system
in a "regular" and a "chaotic" subspace, with respective projectors
$\Pop_r$ and $\Pop_c$; and that the Hamiltonian $\Hop$ may be
written as
\begin{eqnarray}
\label{split}
\Hop\;&=&\;\Hopr\;+\;\Hop_c\;+\;\Vop\;+\;\Vop^{\dagger}\;,
\end{eqnarray}
where $\Hopr=\Pop_r\Hop\Pop_r$ is a "regular" Hamiltonian,
$\Hop_c=\Pop_c\Hop \Pop_c$ is a "chaotic" Hamiltonian, and
$\Vop=\Pop_r\Hop\Pop_c$ couples regular and chaotic states. In the
case of maps, the Hamiltonian formalism has to be recovered by means
of Floquet theory. Eigenvalues of Floquet Hamiltonians fall in
Floquet zones, and for a system driven with a period $T$ in time,
the width of a zone is $2\pi\hbar/T$. In our case we assume $T=1$
and so we identify the 1st Floquet zone with the interval
$[0,2\pi\hbar)$. For the Hamiltonian $\Hop_c$ we assume in our case
(where $a\neq 0$ is always understood)
a continuous spectrum \footnote{One might more
conventionally model $\Hop_c$ by a Random Matrix, of rank
$\propto\hbar^{-1}$, and still reach the crucial result
(\ref{fermi1}), at the price of using some "quasi-continuum" ansatz
at some point. In other words one can assume that  $\Hop_c$ is a
Random Matrix from a Circular  Ensemble, where the density of the
quasienergies is uniform and the eigenstates are statistically
independent of the eigenvalues. In the end the limit of an infinite
dimensional matrix is taken.  We prefer not to repress continuity of
the spectrum, which is a crucial feature of the QAM problem.} As
this Hamiltonian is assumed to be "chaotic", we further assume that
its quasi-energies are non-degenerate and completely fill each
Floquet zone, because typical random matrices drawn from  circular
ensembles have simple spectra, with eigenphases uniformly
distributed in $[0,2\pi]$. The discrete eigenvalues $E_n$ of the
regular Hamiltonian $\Hopr$ are therefore immersed in the continuous
spectrum of the chaotic Hamiltonian $\Hop_c$, so the coupling
perturbation $\Vop$ drives them off the real axis, and the imaginary
parts $-i\hbar\Gamma_n/2$ they acquire are estimated by Fermi's
Golden Rule :
\begin{eqnarray}
\label{fermi} \Gamma_n\;&\sim&\;\frac{2\pi}{\hbar}|\langle
E,c|\Vop^{\dagger}|E_n\rangle|_{E=E_n}^2\;.
\end{eqnarray}
where  $|E,c\rangle$ is the eigenvector of $\Hop_c$ associated with
an eigenvalue $E$ in the continuous spectrum ($\delta-$function
normalization in energy is assumed for such eigenvectors). Choosing
$E_n$ in the 1st Floquet zone of $\Hop_c$, (\ref{fermi}) may be
rewritten as
$$
\Gamma_n\;\sim\frac{2\pi}{\hbar}\|\Pop_0\Vop^{\dagger}|E_n\rangle\|^2
\mu(E_n)\;.
$$
where $\Pop_0$ is projection onto the 1st zone of $\Hop_c$. The
function $\mu(E)$ is the Local Density of States, normalized to $1$,
 of the
vector $\Pop_0\Vop^{\dagger}|E_n\rangle$ with respect to the chaotic
Hamiltonian $\Hop_c$. It yields the probability that a transition
prompted by $\Vop^{\dagger}$ from state $|E_n\rangle$ will lead to
the continuum state $|E,c\rangle$ in the 1st zone. One may introduce
the "ergodic" assumption, that all in-zone transitions have the same
probability. Then
\begin{equation}
\label{fermi1}
\Gamma_n\;\sim\;\frac{1}{\hbar^2}\|\Pop_0\Vop^{\dagger}|E_n\rangle\|^2\;.
\end{equation}
Whether or not the ergodic assumption is accepted, (\ref{fermi1})
may be assumed to hold up to a factor of
order $1$. \\

\subsection{Resonance-assisted decay.} \label{RAD} \noindent  At
1st order in $\epsilon$, the Hamiltonian $\Hopr$ should correspond
to the classical resonant Hamiltonian, and its eigenstates to
quantized tori thereof. Therefore, the coupling $\Vop$ only reflects
classical corrections of higher order than 1st, because  the
resonant Hamiltonian has no coupling between the inside and the
outside of an island. Higher order corrections are present in
$\Hopr$ as well, and adding their  secular (averaged) parts to the
resonant Hamiltonian a new Hamiltonian $\hres$ is obtained, which is
still completely integrable, and shares the action variable $\Ir$ of
the integrable WS pendulum flow (inside the WS separatrix).
Semiclassical quantization of $\hres$ yields energy levels
$e_n=\hres(\Ir_n)$ with quantized actions $\Ir_n=(n+1/2)\hbar$.
\\
In the following, by "perturbation" we mean what is left of
higher-order corrections, after removing averages. Thus the
"unperturbed" quasi-energy eigenvalues of $\Hop_r$ are given by
$e_n+2\pi N\hbar$ with $N$ any integer ( $2\pi\hbar$
is the width of a Floquet zone). In the classical case, the
destabilizing effects of the perturbation are mainly due to
nonlinear resonances. A classical $r:s$ ($r,s$ integers) resonance
occurs at ${\Ir}={\Ir}_{rs}$ if $r\omega({\Ir}_{rs})=2\pi s$, where
$\omega({\Ir})=\partial\hres({\Ir})/\partial{\Ir}$ is the angular
frequency of the integrable motion in the island. The quantum
fingerprints of classical nonlinear resonances are
quasi-degeneracies in the quasi-energy spectrum. A degeneracy
appears in the unperturbed quasi-energy spectrum of $\Hop_r$,
anytime two or more energy levels $e_n,e_m,...$  are spaced by
multiples of $2\pi\hbar$, that is the width of a Floquet zone. In
the vicinity of a classical resonance $\Ir_{rs}$, this is
approximately true, whenever $n-m$ is an integer multiple of $r$;
for, in fact, $|e_n-e_m|\approx \hbar|n-m|\omega(\Ir_{rs})=2\pi\hbar
s|n-m|/r$. Thus, in the semiclassical regime, a classical resonance
induces quantum quasi-degeneracies, which involve whole sequences of
quantized tori. As such tori are strongly coupled by the
perturbation, decay is enhanced. This
quantum effect is Resonance Assisted Decay \cite{BSU02}. \\
The perturbative approach assumes that the sought for  metastable
states basically consist of superpositions of such strongly coupled,
quasi-degenerate states. This leads to replacing $\Pop_r$ in
(\ref{split}) by projection $\Poprr$ onto a quasi-degenerate
subspace,  similar to methods used in \cite{naama}; and so, in order
to use Fermi's rule (\ref{fermi1}), $\Poprr\Hop\Poprr$ must be
diagonalized, and $\Poprr\Vop$ must be specified. To this end we
first write the matrix of $\Poprr\Hop\Poprr$ in a basis of
quasi-degenerate states. Let ${\Ir}_{n_0}$ be a quantized action
close to ${\Ir}_{rs}$. The energy levels $e_{n_0+Nr}$, where the
integer $N$ takes both negative and positive values,  have an
approximately constant spacing, close to $2\pi\hbar s$, and thus
form a ladder. There is one such ladder for each choice of the
integer $n_0$ in the set of the $r$ closest integers to
$n_{rs}\equiv\Ir_{rs}/\hbar-1/2$ (not necessarily an integer), and
we fix one of them. The quasi-energy levels $e_{n_0+Nr}-2\pi\hbar
Ns$ are quasi-degenerate . Denoting $|N\rangle$ the corresponding
eigenstates, the projector $\Poprr$ onto the ladder subspace is a
finite sum $\sum_N |N\rangle\langle N|$, where $N$ ranges between a
$N_*<0$ and a $N^*>0$. The number of terms in the sum is  equal to
the number of nearly resonant states inside the island and so is
given by $L+1$ where $L=N^*-N_*$.  Using that the actions ${I}_N$ of
levels in a $r:s$ ladder are approximately spaced by multiples of
$r\hbar$, $L$ is estimated by
\begin{equation}\label{area}
L\;\approx\;\mbox{\rm Int}\left[\left(\frac{\cal
A}{2\pi\hbar}-1\right)\frac 1r\right]\;.
\end{equation}
If only nearest-neighbor transitions are considered,  the matrix of
$\Poprr \Hop\Poprr$ over the basis $|N\rangle$ is tridiagonal, of
size $L+1$. The off-diagonal elements $v(N)= \langle
N|\Hop|N+1\rangle$ may be semiclassically assumed to slowly change
with $N$, and will be hence denoted simply by $v$. The diagonal
elements are the nearly degenerate quasi-energies
$W(N)=e_{n_0+Nr}-Nsh$, and Taylor expansion of $\hres({\Ir})$ to 2nd
order near ${\Ir}_{rs}$ yields
\begin{equation}
\label{det} W(N)\approx \hres(\Ir_{rs})+2\pi\hbar\frac sr
(n_0-n_{rs}) +\frac{\hbar^2}{2M}(n_0+rN-n_{rs})^2
\end{equation}
where $M=1/\omega'({\Ir}_{rs})$. It follows that, apart from a
constant (independent on $N$) shift, the diagonal matrix elements of
$\Poprr\Hop\Poprr$ are
\begin{equation}
\label{detu}
 W(N)\approx \frac{\hbar^2}{2M}(rN+\delta n)^2\;.
\end{equation}
where $\delta n=n_0-n_{rs}$. Replacing this in $\Poprr\Hop\Poprr$,
one easily recognizes that the classical limit $\hbar\to 0$,
$L\to\infty$ of $\Poprr\Hop\Poprr$ is the classical pendulum
Hamiltonian:
\begin{equation}
\label{classres} {\cal H}_{rs}(\Ir^*,\theta^*)=\frac{r^2
\Ir^{*2}}{2M}+2v\cos (\theta^*)\;,
\end{equation}
in appropriate canonical variables $\Ir^*,\theta^*$
($N\hbar\sim\Ir^*=(\Ir-\Ir_{rs})/r$ as $\hbar\to 0$). This is the
well-known pendulum approximation near a classical resonance
\cite{LL92}, directly derived from quantum dynamics \cite{ZA}, and
is related by a simple canonical transformation to the slightly
different Hamiltonian (\ref{pendapp})
which is used in \cite{BSU02}.\\
The coupling to continuum $\Poprr\Vop$ remains to be specified. Of
all quantized tori in the chain, the closest to the chaotic sea
corresponds to state $|N^*\rangle$, and we assume that  this one
state (in the given chain) is coupled to the continuum. This
assumption implies :
\begin{equation}
\label{lasthop}
 \Vop^{\dagger}\Poprr\;=\; |\chi\rangle\langle N^*|\;,
\end{equation}
where $|\chi\rangle$ is some vector in the chaotic subspace, about
which our one assumption is that it lies in the 1st Floquet zone.
Its norm has the meaning of a hopping amplitude from the "gateway
state" $|N^*\rangle$ to the normalized state
$||\chi||^{-1}|\chi\rangle$. The latter state may be thought of as a
"last beyond the last" nearly resonant state, corresponding to an
unperturbed torus which was sunk into the stochastic sea by the
perturbation and so cannot any more support a regular quasi-mode of
$\Hop_r$. Thus one may denote
$||\chi\rangle||=v(N^*)$, and extrapolate to this last transition,
too, the semiclassical assumption $v(N^*)\sim v$. After all such
additional constructions, the "ergodic assumption" (end of
section \ref{pheno}) is just that this last state has equal
projections on all eigenstates of $\Hop_c$.
Fermi's  rule (\ref{fermi1}) now yields,
for the decay from an eigenstate $|E_m\rangle$ $(0\leq k\leq L)$ of
$\Poprr\Hop\Poprr$,
\begin{equation}
\label{fermix}
 \Gamma_m\sim\frac{v^2}{\hbar^2}|\langle N^*|E_m\rangle|^2\;.
\end{equation}
The labeling $m=0,\ldots,L$ of the eigenstates of $\Poprr\Hop\Poprr$
is arbitrary for the time being. The tridiagonal Hamiltonian is
defined on a chain of states and the scalar product in the last
formula is the value of the eigenfunction $\langle N|E_m\rangle$ at
the rightmost site $N=N^*$ in the chain. Assuming the eigenfunction
to attain its maximum modulus at a site $N_m$, its value at  site
$N^*$ should be of order $\exp (-\xi_m |N^*-N_m|)$. The quantity
$\xi_m^{-1}$ is the fall-off distance of this eigenfunction. For a
tridiagonal Hamiltonian on a chain, arguments by Herbert, Jones, and
Thouless \cite{Thou72}, estimate this distance , as
\begin{equation}
\label{ullmo0}
 \xi_m \sim \ln \left(\frac { {\dk} } { \overline v
}\right )\;,
\end{equation}
where ${{\dk}}$ is the geometric average of the  differences
$|E_j-E_m|$ ($m$ fixed, $j$ variable, $j\neq m$), that is
\begin{equation}
\label{geav} \ln({{\dk}})=\frac{1}{L}\sum\limits_{j(\neq
m)=0}^{L}\ln(|E_j-E_m|)\;,
\end{equation}
and $\overline v$ is the geometric average of the hopping
coefficients $v(N)$, ($N_*\leq N\leq N^*-1$). Under the assumption
$v(N)\approx v=$const., ${\overline v}=v$, and so finally
\begin{equation}
\label{ullmo}
\Gamma_m\;\sim\;\frac{v^2}{\hbar^2}\left(\frac{v}{{{\dk}}}\right)^{2(N^*-N_m)}\;.
\end{equation}
It should be emphasized that the $E_m$ used in (\ref{geav}) are the
eigenvalues of $\Poprr\Hop\Poprr$ and not its diagonal elements
$W_m$ (\ref{detu}),
and that eqn. (\ref{ullmo0}) is not perturbative.

\subsection{Single-Resonance Assisted Decay.}\label{SRAD}

\noindent At any $v\neq 0$, an island hosts a dense set of
resonances, but only a few of them are quantally resolved. On the
other hand, quasi-resonant ladders of states  may be formed, only if
the resonant transitions are not so broad as to involve off-ladder
states. That means that the classical chain of islands should not
be too wide, because its width is determined by the same parameter
$v$ which yields the hopping amplitude between nearly resonant states.\\
In this section we consider the case when a single, not too wide
resonance dominates all the others, so that each metastable state may
be assumed to sit on one of the ladders which are associated with
that resonance. Decay rates $\Gamma_m$ are not affixed to
unperturbed tori, but to eigenstates $|E_m\rangle$ of a ladder
Hamiltonian. For $v=0$ these correspond to quantized tori in the
quasi-resonant ladder, via (\ref{detu}), and their labeling by
$m=0,\ldots,L$  may be chosen accordingly.  For not  quite
small $v$, however, the correspondence between metastable states and
unperturbed tori may be broken, due to avoided crossings; and so
labeling by the original quantum number may not anymore reflect how
deep in the island an eigenstate is located.
\\
We restrict the following discussion to decay "from the center of
the island", meaning that we consider a metastable state, which is
mostly supported in the innermost part of the island, and is labeled
by $m=0$. For this state, (\ref{ullmo}) reads:
\begin{equation}
\label{ullmo1}
\Gamma_0\;\sim\;\frac{v^{2(L+1)}}{\hbar^2D_0^{2L}}\;=\;
 \frac{v^2}{\hbar^2}e^{-2\xi_0
 L}\;\;\;;\;\;\;\xi_0=\ln(\dz)-\ln({v}).
\end{equation}
For quite small $v$, explicit calculation is possible, using for
$E_j$ their unperturbed ($v=0$) values given by (\ref{detu}). This leads to
\begin{equation}
\label{u1n} |E_j-E_0|\approx \hbar^2|r^2j^2-2n_{rs} rj|/2M,
\end{equation}
and then (\ref{ullmo}) with (\ref{ullmo0}) and (\ref{geav}) leads
to
\begin{equation}
\label{u2n}
\xi_0\simeq\ln\left(\frac{\hbar^2}{2Mv}\right)+\frac
1L\sum\limits_{j=1}^{L}\ln(|r^2j^2-2n_{rs} rj|)\;.
\end{equation}
to be used  in (\ref{ullmo1}).
In the average, (\ref{ullmo1})  takes the form (cp.(\ref{area})):
\begin{equation}
\label{conts} \Gamma_0\sim\frac{v^2}{\hbar^2}e^{-\xi_0{\cal A}/\pi
r\hbar}\;,\;.
\end{equation}
This equation shows that the average dependence of $\Gamma_0$ on
$1/\hbar$ is exponential, because the dependence of $\xi_0$ on
$\hbar$ is semiclassically weak. In fact, $v$ is a classical
quantity, and in the limit $\hbar\to 0$ the average over levels
$E_j$ which enters eqn. (\ref{geav}) turns into a purely classical
quantity, given by the phase-space average of $\ln({\cal
H}_{rs}-E_0)$ (cp.(\ref{classres})) over the island.
\begin{figure}[th]
%\includegraphics[width=15cm,angle=0]{gama1}
\includegraphics[width=15cm,angle=0]{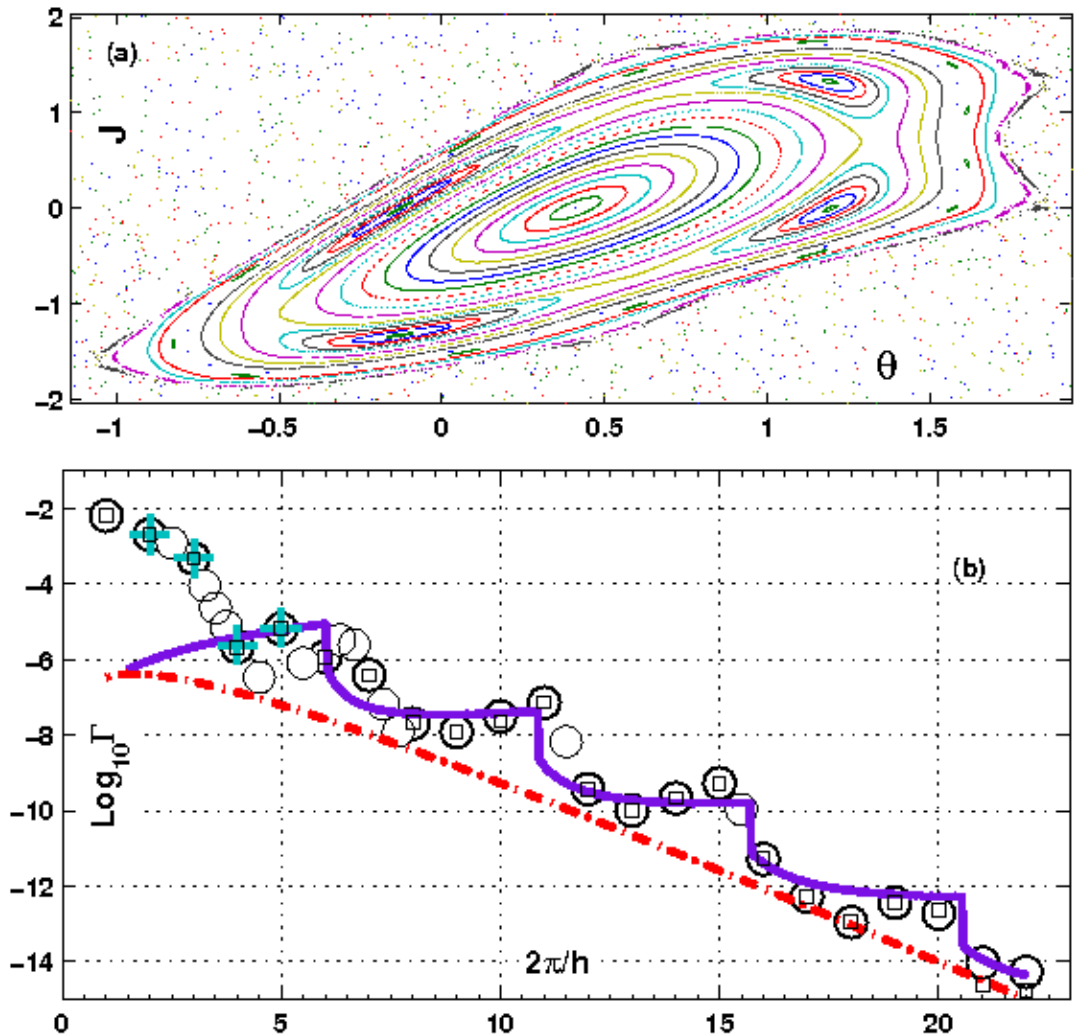}
\caption{(Color online).(a) Phase portrait of map (\ref{map}) with negative sign,
for ${ \tilde k }=2.5$ and $2\protect\pi \Omega =1$. The 4-islands
chain is due to a $4:1$ resonance at $\Ir_{41}\approx 0.43$, with
$v\approx 7.275.10^{-4}$ and $M\approx 3.866$. (b) Minimal decay rate
from
the island shown in (a), calculated by different
numerical (symbols) and theoretical (lines) methods. Circles:
truncated basis method. Pluses: long-time decay of an initial
coherent state in the center of the island. Small Squares : Complex
Scaling Method. Solid line : formula (\ref{ullmo1}), with
(\ref{u2n}). Dotted-dashed line: continuum approximation
(\ref{conts}),(\ref{contsh}).}
%$v_{r:s}=7.275\cdot 10^{-4}$ $%
%m_{r:s}=3.866$ and $J_{r:s}=0.43$ where found from $S^{+}=3.3018$ $%
%S^{-}=2.1018$ and $M_{r:s}=2.5104$.
\label{gama1}
\end{figure}
Approximating the sum in (\ref{u2n}) by an integral, and denoting
${\cal A}_{rs}= 2\pi {\Ir}_{rs}=2\pi\hbar n_{rs}$ the area
enclosed by the $r:s$ resonant unperturbed torus, and $x={\cal
A}_{rs}/{\cal A}$,
\begin{equation}
\label{contsh} \xi_0\sim -\ln\left(\frac{8\pi^2Mv}{{\cal
A}^2}\right)-2+2x\ln(2x)+(1-2x)\ln(|1-2x|)\;\;.
\end{equation}
In formula (\ref{ullmo1}) (with (\ref{u2n})),  $L$ is anyway a discrete
variable, which discontinuously jumps by $1$ anytime increase of
$1/\hbar$ grants accommodation of a new quasi-resonant torus in the
island (cp.(\ref{area})). This produces a stepwise dependence of
$\Gamma_0$ on $1/\hbar$, superimposed on the average exponential
dependence. This structure is smoothed in (\ref{conts}) with
(\ref{contsh}).

\subsection{Numerical results for resonance assisted decay.}

\begin{figure}[ht]
%\includegraphics[width=8cm,angle=0]{continuum.eps}
\includegraphics[width=8cm,angle=0]{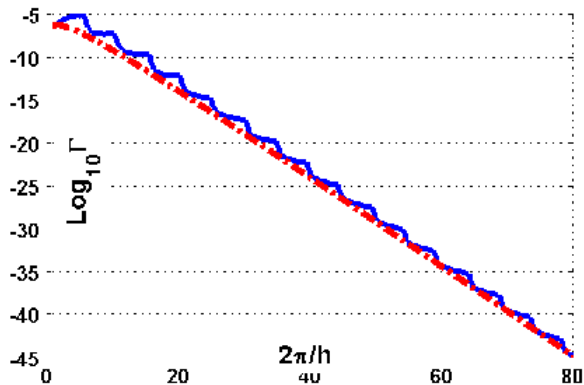}
\caption{(Color online).Comparison between  (\ref{ullmo1}), with (\ref{u2n}), and
the continuum formula (\ref{conts}),(\ref{contsh})}
\label{continuum}
\end{figure}

\begin{figure}[ht]
%\includegraphics[width=8cm,angle=0]{Gamma3Rez.eps}
\includegraphics[width=8cm,angle=0]{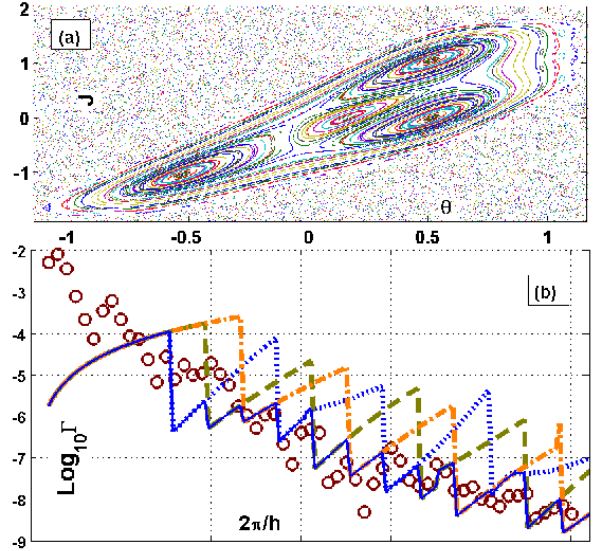}
\caption{(Color online).Illustrating a case with a very broad resonant chain. (a):
Phase portrait of Map(\ref{map}) with negative sign, for ${\tilde
k}=\pi$ and $2\pi\Omega=0.5$. The numerical values for the $3:1$
resonance are $I_{31}=0.13,v=1.9\cdot10^{-3},M=1.52$ (b): Minimal
decay rate from the island, vs $2\pi/\hbar$. Circles: numerical data
(truncated basis). Dotted, dashed, and dash-dotted lines show decay
rates for the three lowest-lying states in the island, computed from
the theory of decay assisted by the large $3:1$ resonance . The
solid line shows the smallest of these.}
 \label{bigres}
\end{figure}

\begin{figure}{}
\includegraphics[width=8cm,angle=0]{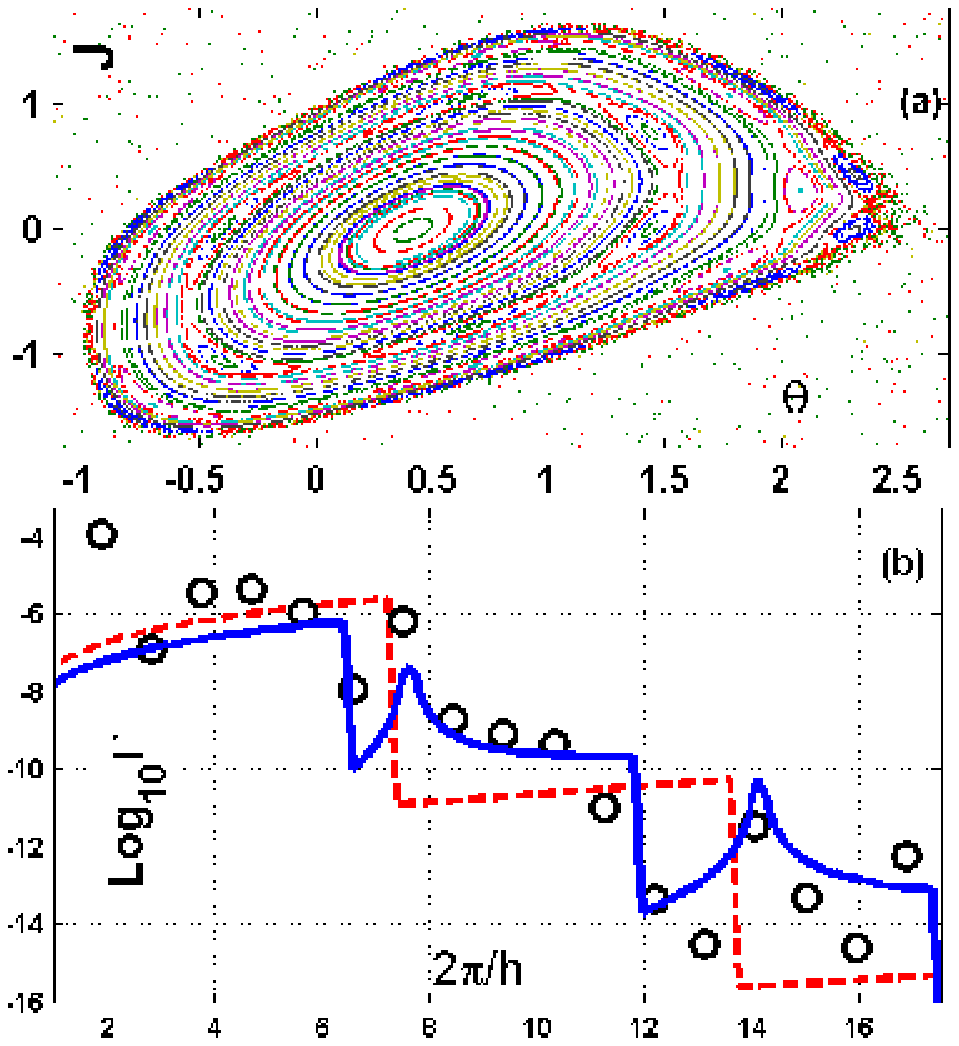}
\caption{(Color online).Illustrating a case with two classical resonances of
comparable size. Here ${\tilde k}=1.329$ and $2\pi\Omega=0.5336$.
Circles in (b) represent the minimal decay rates from  the
truncated basis method. Lines in (b) represent the theory of
Resonance Assisted decay, using either the $6:1$ resonance (solid
line) or the $7:1$ resonance (dashed line). The former resonance
has $v=1.8\cdot 10^{-4}, M=4.504, \Ir_{61}=0.46$, and the latter
has $v=3.1\cdot 10^{-4}, M=2.626, \Ir_{71}=0.93$.} \label{2res}
\end{figure}

\noindent Eqns.(\ref{ullmo1}) and (\ref{conts}) (supplemented by
(\ref{u2n}) or (\ref{contsh})) are the main results of the theory of
resonance assisted decay and are tested against numerical results in
this section. They require specification of $M, v, \Ir_{rs}$ as
input parameters. In the case of well-pronounced chains of resonant
islands, the values of $M, v$, and $\Ir_{rs}$ may be found by
"measuring" areas in the classical phase-space portrait, taking
advantage of the fact that a resonant chain is bounded in between
the separatrices of a pendulum (\ref{classres}). This method was
introduced  in \cite{BSU02},
and is reviewed in Appendix \ref{phspres} for the reader's convenience.\\
For small $v$, unperturbed eigenvalues (\ref{detu}) may be used in
(\ref{geav}) for the purpose of calculating $\dk$, leading to
(\ref{ullmo1}) (with (\ref{u2n})), but one has to beware of values of
$\hbar$, that enforce degeneracy of the unperturbed spectrum
(\ref{detu}). Inadvertent use of the spectrum (\ref{detu}) in
(\ref{geav}) in such cases leads to artificial divergence of
$\Gamma$. In the actual spectrum to be used in (\ref{geav}), degeneracies
are replaced by avoided crossings, which may lead to local enhancement
of resonance-assisted tunneling, as discussed in  Appendix \ref{degeneracy}
and shown in Fig. \ref{deg}.
\\
A case with a single dominant resonance $4:1$ is presented in
Fig.\ref{gama1} . Decay rates calculated from (\ref{ullmo1}) with
(\ref{detu}) and (\ref{u2n}) are shown in (b) for the innermost
state in the island which is shown in Fig.\ref{gama1}(a). Also shown
are results of directly calculating $\Gamma$s, by methods described
in Appendix \ref{metastates}. Formula (\ref{ullmo1}), with
(\ref{u2n}), is seen to correctly reproduce the actual $\Gamma$s, in
order of magnitude at least. The stepwise dependence predicted in
\cite{BSU02} and explained in the end of previous section is here
remarkably evident. The continuum (quasiclassical) approximation
(\ref{conts}),(\ref{contsh}) is also shown in the same Figure. In
Fig.\ref{continuum} it is seen to better and better agree
with  (\ref{ullmo1}), with (\ref{u2n}), at smaller and smaller values of $\hbar$.\\
As remarked in section \ref{SRAD}, the presented theory, being
essentially perturbative,  is expected to fail in the case of large
classical resonances. An example is presented in Fig.\ref{bigres}.
Finally in Fig.\ref{2res} we present a case with two resonant chains of
comparable size. Results are not well described by the theory based
on either resonance, and thus  appear to contradict a somewhat
natural expectation, that each resonance should contribute its own
set of metastable states. The  single-ladder picture may not be
adequate in such cases, which therefore remain outside the present
scope of the theory.

\section{Discussion and Conclusions}

The decay rates of some metastable states related to phase space
islands were calculated and  the required theoretical framework was
developed. The main results of the paper are (\ref{gwsn}),
(\ref{ullmo}) and (\ref{ullmo1}), where the decay rates of wave
packets in phase space islands were calculated for various
conditions. If the effective Planck's  constant is sufficiently
large, so that island chains cannot be resolved on its scale,
standard WKB theory was found to work well. As the effective
Planck's constant is decreased, island chains are resolved, and
dominate the decay by the mechanism of resonance assisted tunneling.
Its signature here is the step structure of Fig.\ref{gama1}(b). The
average slope of $\ln \Gamma$ (see (\ref{conts})) as a function of
$1/\hbar$ is $-\xi_0 {\cal A}/\pi r$, with $\xi_0$ given by
(\ref{contsh}) and is independent of $\hbar$.  In the WKB regime a
different slope is found (see (\ref{gwsn})). For some values of the
parameters, resonance-assisted tunneling can be further enhanced by a degeneracy
between a semiclassical state deep inside the island and one that is
close to the boundary. An interesting question is about possible
effects of this kind,  due to "vague tori" \cite{SR82}, i.e., to classical structures
which quantally act as tori, in spite of lying outside an island.
Indeed, observations in \cite{FGR00} have suggested a possible role for
cantori in enhancing QAMs at times.\\
The theory strongly relies on the dominance of {\it one}  resonant
island chain
If the phase space area
of the  chain is not small as is the case in Fig.\ref{bigres} or
if the tunneling is assisted by two (or more) resonant island
chains of approximately equal strength, as is the case in Fig.\ref{2res},
our theory requires modification.

The theory is relevant for systems of experimental interest but
the steps of Fig.\ref{gama1} were found for a regime where the
decay rate is too small to be experimentally accessible.
Overcoming this problem is a great theoretical and experimental
challenge.

\begin{acknowledgments}
It is a pleasure to thank  P.Schlagheck, D.Ullmo,
 E.E.Narimanov, A.B\"acker, R.Ketzmerick, N.
Moiseyev and J.E. Avron
for useful discussions and correspondence. This research was supported in part by the
Shlomo Kaplansky Academic Chair, by the US-Israel Binational Science Foundation (BSF),
by the Israeli Science Foundation (ISF), and by the Minerva Center of Nonlinear
Physics of Complex Systems. L .R. and I.G. acknowledge partial support from the MIUR-PRIN
project "Order and chaos in extended nonlinear Systems: coherent structures, weak
stochasticity, and anomalous transport".
\end{acknowledgments}

\appendix
\section{Metastable states.} \label{metastates}

\noindent Our methods of computing decay rates $\Gamma$  were: (1)
basis truncation, (2) complex scaling, (3) simulation of wave-packet
dynamics. In the cases investigated in this paper, the most
economical one, and thus the one of our prevalent use, was (1). The
other two methods were used to cross-check results of (1) in a
number of cases. The observed agreement between such completely
independent computational schemes demonstrates the existence of
resonances (in the sense of metastable states).

\subsection{Basis Truncation.}

\begin{figure}{}
\includegraphics[width=8cm,angle=0]{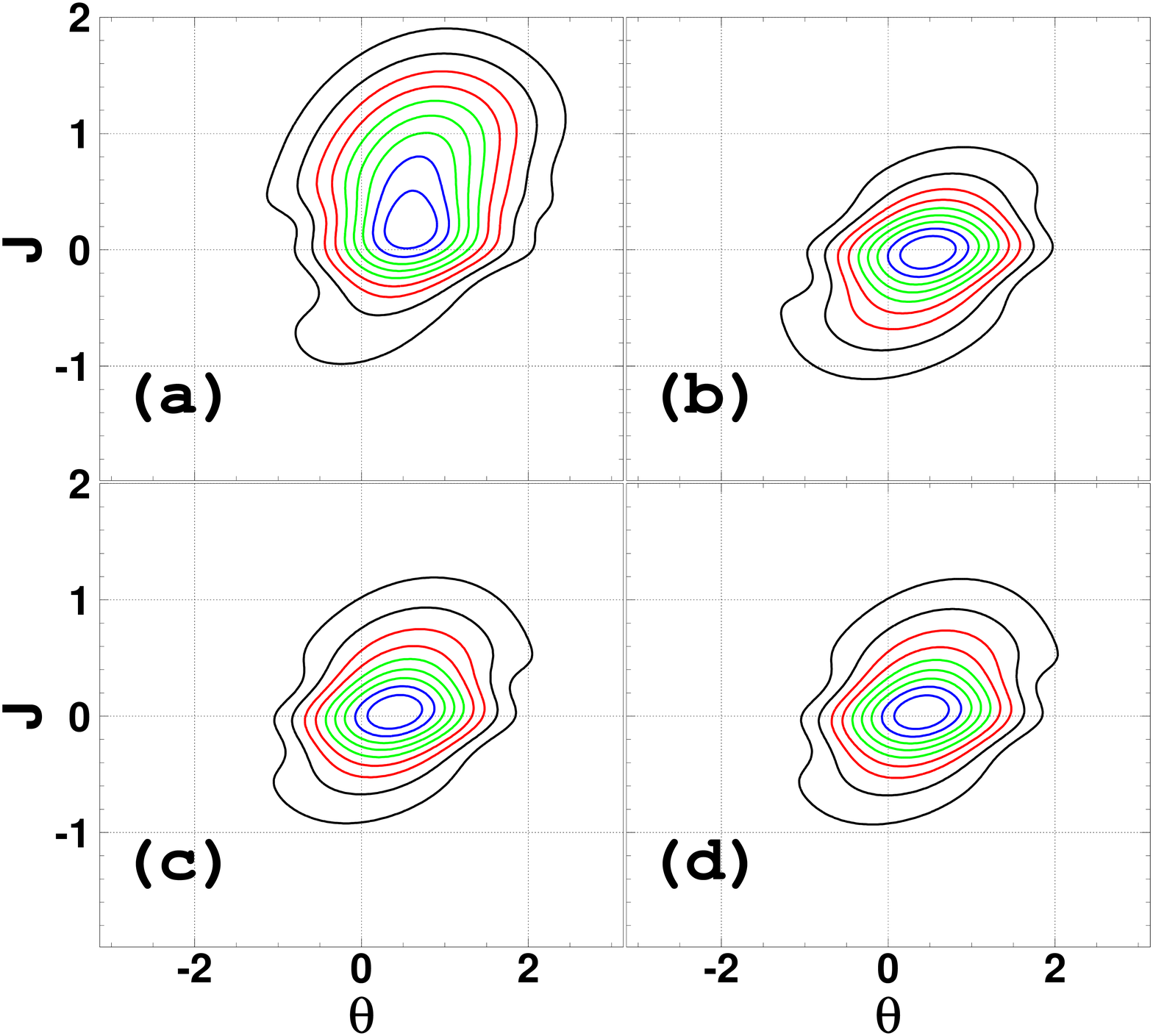}
\caption{(Color online).(a),(b),(c): Evolution of an initial coherent 
state located
at the center of the island of Fig.\ref{gama1}, for ${\tilde
k}=2.5$, $2\pi\Omega=1$, $\hbar=0.25$, numerically simulated by a
FFT method with basis size $2^{17}$. (a),(b),(c) show contour plots
of the Husimi functions at times t=100, t=1000, t=16000
respectively. (d): Husimi contour plot of the eigenfunction of the
truncated basis dynamics, which has the largest overlap with the
initial coherent state. The size of the truncated basis is $4096$.
}\label{whatsleft}
\end{figure}

\begin{figure}{}
\includegraphics[width=8cm,angle=0]{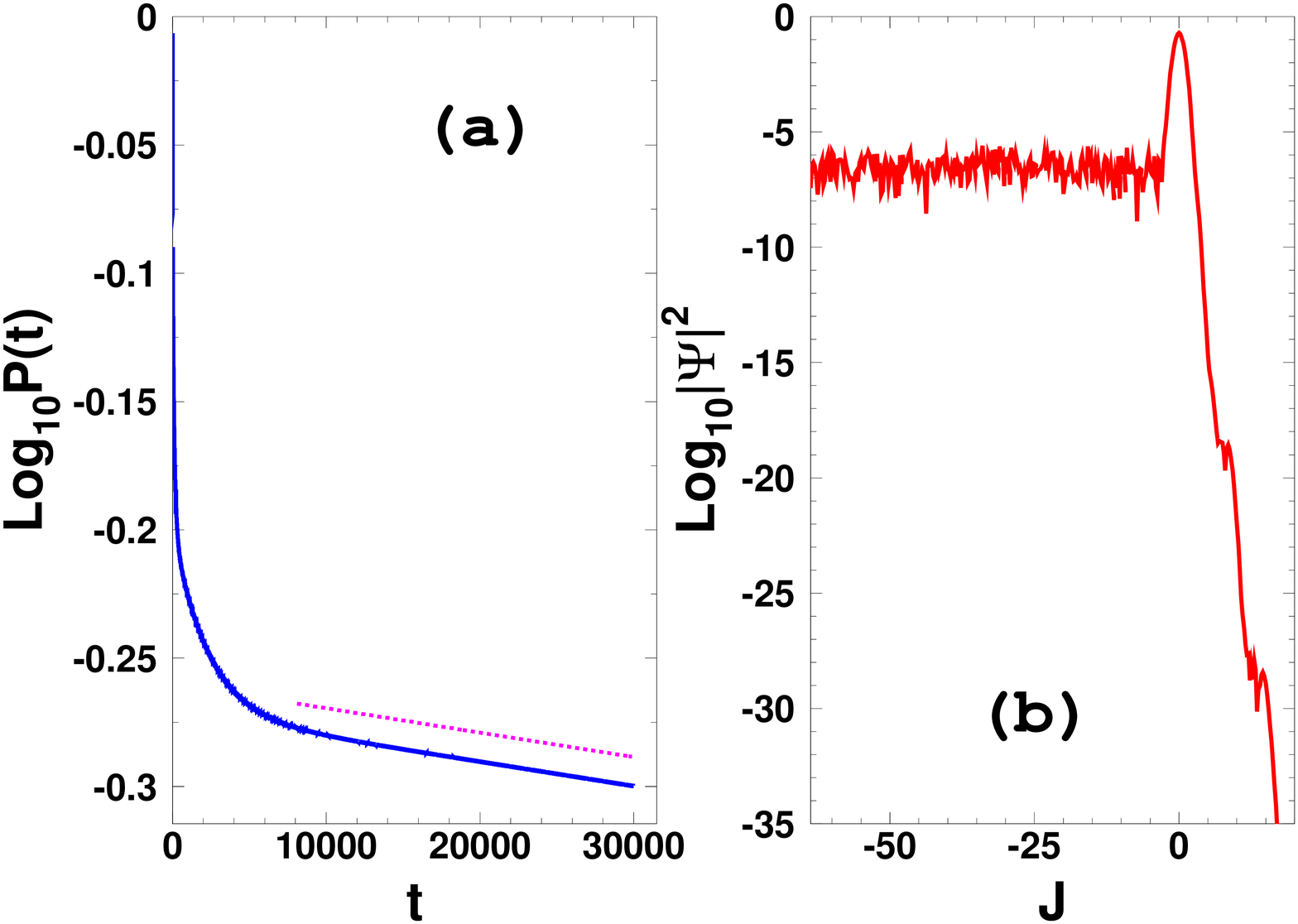}
\caption{(Color online).(a): Decay in time of the probability 
in a momentum window
containing the island, for the same parameter values as in
Fig.\ref{whatsleft}, and for the same choice of the initial state.
The size of the basis used in the simulation is $2^{18}$. The
straight line has slope $\approx-5.66$. (b) squared modulus of the
truncated basis eigenfunction presented in Fig.\ref{whatsleft}(d),
in the momentum representation.}\label{eigf+pt}
\end{figure}

\noindent Let $\Uop$ denote the unitary evolution operator that is
obtained by quantizing map (\ref{map}), as described in
sect.\ref{quant}; and let $|n\rangle$, $(n\in\ZM)$ be the eigenvectors
of the angular momentum operator ${\hat{J}}$, such that
$\langle\theta|n\rangle=(2\pi)^{-1/2}\exp(in\theta)$. "Basis
truncation" consists in replacing $\Uop$ by
$\Uop_{\nu}=\Pop_{\nu}\Uop\Pop_{\nu}$, where $\Pop_{\nu}=\sum_{
|n|\leq \nu}|n\rangle\langle n|$. This  introduces an artificial
dissipation, which turns the quantum dynamics from unitary to
sub-unitary. The eigenvalues $z_j$, $(j=1,\ldots, 2\nu+1)$ of
$\Uop_{\nu}$ lie inside the unit circle, with positive decay rates
$\Gamma_j=\ln(1/|z_j|)/2$. As $\nu$ is increased, most of them move
towards the unit circle, but some appear to stabilize at fixed
locations inside the circle, because they approximate actual,
subunitary eigenvalues of the exact $(\nu=\infty)$ non-dissipative
dynamics. The seeming contradiction to unitarity of the limit
dynamics $\Uop$ is solved by the observation that the eigenfunctions
of $\Uop_{\nu}$, which are associated with such eigenvalues,  tend
to {\it increase} in the negative momentum direction
(Fig.\ref{eigf+pt}), as expected of Gamov states, and if this
behavior is extrapolated to the limit, then they cannot belong in
the Hilbert space wherein $\Uop$ acts unitarily. In order to make
room for such non-unimodular eigenvalues,  $\Uop$ must be extended
to a larger functional space. Complex scaling, to be described in
the next subsection, provides a consistent method of doing
that.\\
Simulations of wave packet dynamics, performed in the total absence
of any dissipation whatsoever, confirm this interpretation of the
stable subunitary eigenvalues (Fig.\ref{whatsleft}). The initial
wave packet is a coherent state supported near the center of the
island . We use a FFT algorithm, so the computed evolution is fully
unitary, and reliably reproduces the exact evolution over a long
time, thanks to the large dimension of the FFT . We compute the
decay in time of the probability in a momentum window which contains
the classical island. After an initial rapid decay, due to escape of
the part of the distribution which initially lies outside the
island,  the decay turns into a clean exponential with rate
$\gamma\approx 5.66$ (Fig.\ref{eigf+pt}). Among the eigenfunctions
of the truncated basis evolution, which correspond to stabilized
eigenvalues, we select the one which has largest overlap with the
chosen initial state. We thus find that (i) the decay rate $\Gamma$
of this eigenfunction is $\approx\gamma$, (ii) the Husimi function
of that part of the wavepacket, which has survived in the chosen
window until the end of the dynamical calculation, nearly reproduces
the Husimi function of the eigenfunction (Fig.\ref{whatsleft}).\\
Numerical computation of long-time exponential decay may be less
easy than in the above particular example. If an initial state is
overlapped by several metastable states with slightly different
$\Gamma$, resolving them may take quite a long computational time,
and hence a huge basis, because of accelerated motion outside the
island.

\subsection{Complex Scaling.}

\noindent
Scattering resonances may be sometimes  computed by
diagonalizing a non-hermitean Hamiltonian, which is constructed by
"complex coordinate" methods such as analytic dilation, and the
like \cite{MSV03,RS78}.
Despite absence of  scattering theory, a method of this sort was devised for the subunitary
eigenvalues considered in this paper, as follows. Any function
$\psi(\theta)$ over $[0,2\pi]$ is at once a function $\tilde\psi(z)$
of the complex variable $z$ running on  the unit circle. Let ${\cal
B}$ denote the class of those functions $\psi(\theta)$, which can be
analytically continued to the whole complex plane, except possibly
the origin. For given $1\geq\rho>0$ the {\it scaling operator} $h_{\rho}$
is defined to act on the functions of this class as in
$(h_{\rho}\psi)(\theta)=\tilde\psi(\rho e^{i\theta})$. The crucial
property of the operator $\Uop$, which makes the present
construction possible, is that of transforming functions in  class
${\cal B}$ in functions in the same class, and so the operator $
U_{\rho}\equiv h_{\rho}\Uop h^{-1}_{\rho} $ is a well defined
operator in ${\cal B}$. This operator trivially extends by
continuity to an operator $U_{\rho}$
 \cite{G06} which is  defined on the whole of $L^2([0,2\pi])$.
This extension formally amounts to defining $\Uop$ also on a class
of functions, which are not square integrable. The new ``functions''
thus acquired in the domain of the evolution operator may be very
singular objects; e.g., in the momentum representation,
 they are  allowed to exponentially diverge at infinity.
In the special case $\hbar=1/n$ with $n$ integer, $\Uop_{\rho}$  has the form :
$$
\Uop_{\rho}=\rho^n e^{i{n\hat\theta}}e^{-i k_+\cos({\hat\theta})}
e^{k_-\sin({\hat\theta})}
e^{-i{\hat J}^2/2}\;\;,\;k_{\pm}=\frac 12{\tilde k}\left|
\rho\pm\rho^{-1}\right|\;.
$$
which restitutes $\Uop$ for $\rho=1$. The eigenvalues of
$\Uop_{\rho}$ as a $L^2$ operator are at once eigenvalues of the
``extended'' $\Uop$, and each of the latter eigenvalues is an  eigenvalue
of $\Uop_{\rho}$ for $\rho$ sufficiently distant from $1$.
 The "complex scaling" method of computing
$\Gamma$, which is mentioned in the main text, consists in
diagonalization of $\Uop_{\rho}$. At small $\hbar$ this method is
computationally problematic, due to exponentially large elements in the
matrix of $\Uop_{\rho}$.

\begin{figure}{}
\includegraphics[width=9cm,angle=0]{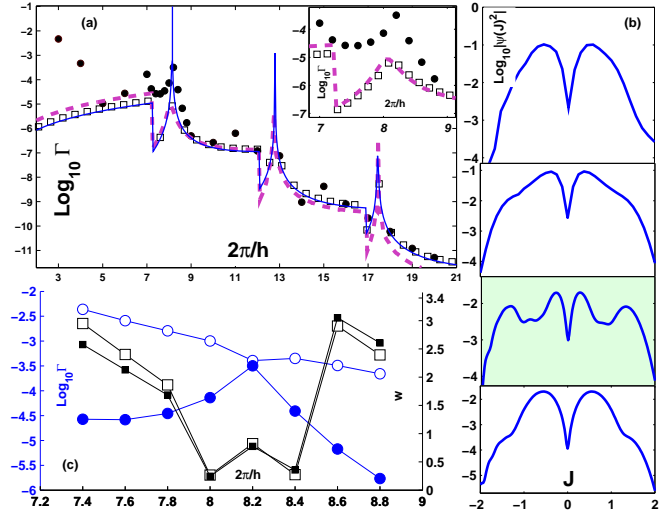}
\caption{(Color online).(a) Decay rate from the 1st 
excited state in the island,
vs $1/\hbar$, for the same parameter values as in Fig.\ref{gama1}.
Dots: numerical (truncated basis method). Solid line: formula
(\ref{ullmo}) with (\ref{geav}), using the unperturbed ladder
spectrum (\ref{detu}). Squares: same formula, using the
perturbed  spectrum of the tridiagonal matrix with (\ref{detu})
as diagonal elements.
Dashed line: same formula, using the actual spectrum, semiclassically reconstructed from
the layout of tori in the island.
(b) the eigenstate of the (truncated) evolution operator,
corresponding to the decay rate shown in (a) for values of $1/\hbar $: $7$, $8$, $8.2$
and $8.6$ (from above down), in the momentum $J$
representation. Only a $J-$interval of roughly the size of the
island is shown. (c) Decay rate $\Gamma$ (circles) and real
quasi-energy $w$ (squares) vs $1/\hbar$, for the 1st (full
symbols) and the 5th (empty symbols) excited states.} \label{deg}
\end{figure}

\section{Inferring Parameters of a Classical Resonance from Phase
Portraits.}\label{phspres}

\noindent Parameters $\Ir_{rs}$, $M$, and $v$ of a classical $r:s$
resonance respectively specify the value of the unperturbed action
where the resonance is located, the inverse nonlinearity
$1/\omega'(\Ir_{rs})$, and the strength of the resonant harmonic
perturbation. These parameters are indispensable for the formalism
described in sect.\ref{RAD}, and may be retrieved  from the
phase-space portrait, using formulae taken from \cite{ES05}. Here we
reproduce  a sketchy derivation for the reader's convenience. Motion
in a $r:s$ resonant chain is approximately described by a pendulum
Hamiltonian \cite{LL92}, which may be written in the form (which is
canonically equivalent to (\ref{classres}):
\begin{equation}\label{pendapp}
H_{rs}({\Ir},\varphi)=\frac{1}{2M}(\Ir-\Ir_{rs})^2\;+\;2v\;\cos(r\varphi)\;.
\end{equation}
In this paper, $\Ir$ is the action variable of the WS-pendulum
Hamiltonian (\ref{ws}). The Separatrices of Hamiltonian
(\ref{pendapp}) are the curves $\Ir=\Ir_{\pm}(\varphi)=\Ir_{rs}\pm
2\sqrt{Mv(1-\cos(r\varphi))}$, and so the phase areas $S_{\pm}$ they
enclose satisfy:
\begin{eqnarray}
\label{phareas}
S_+\;+\;S_-\;&=&\;4\pi \Ir_{rs}\;,\nonumber\\
S_+\;-\;S_-\;&=&2\int_0^{2\pi}d\varphi\;(\Ir_+(\varphi)-\Ir_{rs})\;=\;16\sqrt{2Mv}\;.
\end{eqnarray}
The monodromy matrix of the stable period-$r$ orbit that is
responsible for the resonant chain is easily obtained by linearizing
the flow (\ref{pendapp}) near the stable equilibrium point(s). Its
trace is found to be ${\cal M}=2\cos(r\nu)$ where $\nu=r\sqrt{2v/M}$ is
the angular frequency of the small pendulum oscillations. This leads
to
\begin{equation}
\label{mono} \sqrt{2v/M}\;=\;r^{-2}\arccos({\cal M}/2)\;.
\end{equation}
The phase-space  areas $S_{\pm}$ and the Monodromy Matrix  can be
numerically determined, and once their values are known
eqs.(\ref{phareas}) and (\ref{mono}) can be solved for $\Ir_{rs}$,
$M$, and $v$.

\section{Avoided Crossings.}\label{degeneracy}

\noindent
An exact degeneracy arises in the unperturbed ladder
spectrum (\ref{detu}) whenever two quantized actions in the ladder
are symmetrically located with respect to the the resonant action
$\Ir_{rs}$. This requires $\delta n=0$ or $\delta n=\pm r/2$ in
(\ref{detu}) and so, if $(n_i+1/2)\hbar$ is the smallest quantized
action in the ladder, such symmetric pairs exist if, and only if,
\begin{equation}
\label{dege}
\frac{1}{\hbar}=\frac{2n_i+lr+1}{2\Ir_{rs}}\;\;\mbox{\rm and}\;\;
L(\hbar)\geq l\;,
\end{equation}
for some integer $l\geq 1$. In the above inequality, the length $L$
of the ladder depends on $\hbar$ as in (\ref{area}). In the case of
Fig.\ref{gama1}, where $n_i=0$ (the ground state), (\ref{dege}) is
never satisfied for $1/\hbar<20$. The data in Fig.\ref{gama1} were computed
using the spectrum (\ref{detu}), and no significant
difference could be found between them and data computed by using the
actual spectrum, semiclassically reconstructed from
the layout of tori in the island.
In Fig.\ref{deg}, where $n_i=1$
(the "1st excited state"),  (\ref{dege}) is satisfied for $1/\hbar=8.2, 12.8,
17.44$, corresponding to $l=1,2,3$. At such values of $1/\hbar$ the spectrum
(\ref{detu})(with the appropriate $\delta n$) is degenerate and using it in
formulas (\ref{geav}),(\ref{ullmo})
obviously
causes $\Gamma$ to diverge,
as shown by the narrow peaks in Fig.\ref{deg}. Such artifacts
disappear on inserting the proper (perturbed) spectrum, obtained by
diagonalizing the ladder Hamiltonian with the diagonal elements
given by (\ref{detu}), because for $v\neq 0$ the spectrum is never
degenerate. Nevertheless avoided crossings take place at the values
(\ref{dege}) of $\hbar$, giving rise to local peaks in the
dependence of $\Gamma$ on
$\hbar$.\\
 At the same values of $1/\hbar$ avoided crossings are observed even
between subunitary, stabilized eigenvalues of the truncated
evolution operators (see App.\ref{metastates}). This is shown in
Fig.\ref{deg}(c). The eigenvalues are written $z=e^{-\Gamma/2-iw}$
and it is seen that, as $1/\hbar$ approaches a value $\approx 8.2$,
a pair of complex eigenvalues undergo a close avoided crossing. The
corresponding states exhibit standard behavior  at avoided
crossings. The distribution in momentum $J$ of one of them is shown
in Fig.\ref{deg}). This state nominally corresponds to the $n=1$
unperturbed state, and in fact in (b)(top) it looks similar to the
1st excited state of a harmonic oscillator. The other state
nominally corresponds to the $n=5$ unperturbed state. At the avoided
crossing (3d inset from top in (b)) the former state significantly
expands over the island, because it is basically a superposition of
two unperturbed states, which are located symmetrically with respect
to $\Ir_{41}$. This gives rise to a local enhancement of the decay
rate.

\end{document}